\documentclass[aps,prl,reprint,amsmath,amssymb,nofootinbib,nobibnotes]{revtex4-2}
\usepackage{bm,microtype}
\usepackage{dcolumn,mathrsfs}
\usepackage{hyperref}
\usepackage{xcolor}
\usepackage{amssymb}
\hypersetup{hidelinks,pdftitle={Separation of gravitational perturbations of Myers-Perry black holes},pdfauthor={Jun-Kun Zhao}}
\allowdisplaybreaks[2]
\newcommand{\dd}{\mathrm d}
\newcommand{\ii}{\mathrm i}

\begin{document}
\title{Separation of gravitational perturbations of Myers--Perry black holes}

\author{\textbf{Jun-Kun Zhao}}
\email{junkunzhao@itp.ac.cn}
\affiliation{Institute of Theoretical Physics, Chinese Academy of Sciences, Beijing 100190, China}
\author{\textbf{Hong-Da Lyu}}
\email{hongdalyu@buaa.edu.cn}
\affiliation{Key Laboratory of Particle Physics and Particle Irradiation (MOE), \\Institute of Frontier and Interdisciplinary Science, \\Shandong University, Qingdao, Shandong 266237, China}

\begin{abstract}
Gravitational perturbations of higher-dimensional rotating black holes generally obey coupled partial differential equations, making their separation a longstanding challenge. We separate linearized gravitational perturbations of nonextremal, asymptotically flat five-dimensional Myers--Perry black holes with two nonzero spins of unequal magnitude and arbitrary azimuthal numbers. The Killing--Yano symmetry induces a first-order angular operator that, together with a commuting second-order operator, determines the angular profiles of all five gauge-invariant curvature components. The joint angular harmonics form a complete basis at real frequencies, and all five components of each separated mode share a single radial amplitude satisfying a second-order scalar equation obtained through the Carter interchange. The corresponding metric perturbations can be reconstructed locally. We further extend this angular structure to Myers--Perry black holes in arbitrary dimensions with nonzero spins of pairwise distinct magnitudes. Second-order symmetry operators associated with the Killing tensors induce mutually commuting angular operators on all curvature polarizations, with coefficients independent of radius. In odd dimensions, a Killing--Yano three-form generates an additional first-order angular operator commuting with this family. This common angular structure provides a route toward complete separation of gravitational perturbations in higher-dimensional rotating black holes.
\end{abstract}

\maketitle

\emph{Introduction.---} The study of perturbations of black holes plays a central role in understanding their dynamics, stability, and gravitational-wave emission~\cite{Berti:2025hly}. The characteristic ringdown oscillations, described by quasinormal modes, underpin black hole spectroscopy and offer a powerful probe of strong-field gravity~\cite{Berti:2005ys,Berti:2009kk}. A central question is how to formulate the perturbation equations so that they separate into radial and angular ordinary differential equations, providing a tractable framework for studying the linear dynamics of both gravitational and matter fields. For Kerr black holes, the extreme Weyl scalars satisfy decoupled, separable equations~\cite{Teukolsky:1972my,Teukolsky:1973ha}, enabling the reconstruction of radiative metric perturbations~\cite{Chrzanowski:1975wv,Wald:1978vm}. These landmark developments laid the foundations of modern black hole perturbation theory~\cite{Berti:2025hly}. More recently, coupled gravitoelectromagnetic perturbations of Kerr--Newman black holes have been separated~\cite{Hintz:2026jiw}, with the separation subsequently validated numerically~\cite{Destounis:2026ljd} and extended to other type-D spacetimes~\cite{Yoshida:2026yly,Zhao:2026gen,Gray:2026rzu}. These developments show that coupled perturbations can retain a rich separable structure.

Despite these advances, separating gravitational perturbations of higher-dimensional rotating black holes remains a major challenge. A paradigmatic example is the Myers--Perry (MP) family~\cite{Myers:1986un,Gibbons:2004js}, the natural generalization of the Kerr black hole, which admits multiple independent rotation parameters, while gravitational waves carry $D(D-3)/2$ physical polarizations in $D$ dimensions. In the nonrotating limit, gravitational perturbations admit decoupled master equations~\cite{Kodama:2003jz}. For the general rotating case, hidden symmetries ensure the separability of geodesic motion and scalar waves~\cite{Frolov:2003en,Frolov:2002xf,Frolov:2006dqt}, but the geometry of the principal null congruences obstructs a direct Teukolsky-type decoupling of the extreme Weyl components~\cite{Durkee:2010qu}. Previous studies have examined selected gravitational modes in the equal-spin case~\cite{Tsuchiya:2020hur} and constructed separated solutions for certain gravitational perturbations~\cite{Lunin:2025yth}. More recently, gravitational perturbations of singly spinning MP black holes were separated for modes preserving the transverse-sphere symmetry~\cite{Santos:2026pbu}. However, the separation of general gravitational perturbations with multiple nonzero, unequal spins has remained an open problem.

In this Letter, we make progress on this problem by deriving separated equations for gravitational perturbations of nonextremal, asymptotically flat five-dimensional MP black holes with two nonzero spins of unequal magnitude. The kernel of the principal tensor singles out a Killing direction, $\xi_K=ab\,\partial_t+b\,\partial_\phi+a\,\partial_\psi$, along which local Kaluza--Klein reduction yields a four-dimensional metric conformal to a Carter-type metric of Petrov type D. The Carter structure revealed by this reduction motivates our choice of coordinates, while the Killing--Yano (KY) three-form dual to the principal tensor provides the symmetry underlying our separation. We show that a first-order angular operator induced by the KY symmetry, together with a commuting second-order operator, determines the relative profiles of all five polarizations and yields a complete angular basis at real frequencies. The Carter interchange then yields a single second-order radial equation, while the linearized Einstein equations select the physical angular--radial pairings. For the resulting solutions, all five polarizations share a single radial amplitude, and the corresponding metric perturbations can be reconstructed locally up to gauge. In the singly spinning limit, our construction recovers Santos's separation equations in the corresponding symmetry sector and provides a KY interpretation of his angular operators~\cite{Santos:2026pbu}. 

We further extend this angular construction to asymptotically flat MP black holes in arbitrary dimensions with nonzero spins of pairwise distinct magnitudes. Second-order symmetry operators associated with the Killing tensors induce mutually commuting angular operators on all $D(D-3)/2$ gauge-invariant curvature components, with coefficients independent of radius. In odd dimensions, a KY three-form generates an additional first-order angular operator commuting with this family. These identities define the same joint angular eigenvalue problem on every radial slice and provide a route toward complete separation of gravitational perturbations in higher-dimensional rotating black holes.

\emph{Model and setup.---}
We consider the five-dimensional, asymptotically flat MP black hole~\cite{Myers:1986un}. In Carter coordinates $(T,r,y,P,w)$ with signature $(-++++)$, its line element reads~\cite{Chen:2006xh}
\begin{equation}\label{eq:metric}
\dd s^2=-\frac{X}{\Sigma}\mathcal A^2
+\Sigma\left(\frac{\dd r^2}{X}+\frac{\dd y^2}{Y}\right)
+\frac{Y}{\Sigma}\mathcal B^2+r^2y^2\mathcal C^2 \,,
\end{equation}
where $\mathcal A=\dd T+y^2\dd P$, $\mathcal B=\dd T-r^2\dd P$, and $\mathcal C=\dd w-(ab/\Sigma)(\mathcal A/r^2+\mathcal B/y^2)$. Here $\Sigma=r^2+y^2$, $X=\Delta/r^2$, $Y=(a^2-y^2)(y^2-b^2)/y^2$, and $\Delta=(r^2+a^2)(r^2+b^2)-\mu r^2$, with $\mu$ the mass parameter and $a,b$ the two rotation parameters. We consider nonextremal black holes with $\mu>(|a|+|b|)^2$. The exterior region is $r>r_+$, where $r_+$ is the largest positive root of $\Delta$.

The Carter coordinates are related to the Boyer--Lindquist coordinates $(t,r,\theta,\phi,\psi)$ by $y^2=a^2\cos^2\theta+b^2\sin^2\theta$, $t=T+(a^2+b^2)P+ab\,w$, $\phi=aP+bw$, and $\psi=bP+aw$~\cite{Chen:2006ea,Chawla:2022ogv}. Here $r$ is unchanged, $0\leq\theta\leq\pi/2$, and both $\phi$ and $\psi$ have period $2\pi$. For unequal spin magnitudes, cases with $ab<0$ are included by reversing $\phi$ or $\psi$. For equal spin magnitudes, $a^2=b^2>0$, $y$ becomes constant and the Carter coordinates degenerate, while the enhanced $U(2)$ symmetry organizes the perturbations into Wigner harmonics~\cite{Murata:2008yx}. In what follows, we focus on gravitational perturbations of MP black holes with unequal spin magnitudes and take $a>b>0$ without loss of generality, so that $b\leq y\leq a$.

For a metric perturbation $h_{ab}$, we seek Fourier solutions of the linearized vacuum equations,
\begin{equation}\label{eq:vacuum}
\delta R_{ab}[h]=0 \,,\quad h_{ab}= e^{-\ii\omega T+\ii k_P P+\ii k_w w} \mathfrak{h}_{ab}(r,y) \,.
\end{equation}
Although $T$ differs from Boyer--Lindquist time $t$, $\partial_T=\partial_t$, so $\omega$ is the same in both coordinate systems. In Boyer--Lindquist coordinates, the Fourier factor is $e^{-\ii\omega t+\ii m_\phi\phi+\ii m_\psi\psi}$, where the $2\pi$ periods of $\phi$ and $\psi$ require $m_\phi,m_\psi\in\mathbb Z$. Using the transformation between Boyer--Lindquist and Carter coordinates and comparing with~\eqref{eq:vacuum}, we obtain $k_P=a m_\phi+b m_\psi-(a^2+b^2)\omega$ and $k_w=b m_\phi+a m_\psi-ab\omega$. Thus, at fixed $\omega$, the Fourier coefficients $k_P$ and $k_w$ are determined by the azimuthal numbers $m_\phi$ and $m_\psi$.

\emph{Curvature variables and polarization.---}
The type-D structure of the MP background allows us to construct gauge-invariant curvature variables for its perturbations~\cite{Pravda:2007ty,Durkee:2010qu}. Following the higher-dimensional Newman--Penrose approach~\cite{Durkee:2010xq}, we choose a null frame $(L,N,E_i)$, with $i=+,-,3$, where $L$ and $N$ are aligned with the two Weyl-aligned null directions. In Carter coordinates $(T,r,y,P,w)$, the frame vectors are $L^a=X^{-1}(r^2,-X,0,1,ab/r^2)$, $N^a=(2\Sigma)^{-1}(r^2,X,0,1,ab/r^2)$, $E_\pm^a=(p_\pm\sqrt{2Y})^{-1}(-y^2,0,\mp\ii Y,1,-ab/y^2)$, and $E_3^a=(ry)^{-1}(0,0,0,0,1)$, where $p_\pm=r\pm\ii y$. They satisfy $L^aN_a=-1$, $E_+^aE_{-a}=1$, and $E_3^aE_{3a}=1$, with $E_i$ forming a complex basis for the three-dimensional space orthogonal to $L$ and $N$.

In this frame, the background Weyl projections $C_{abcd}L^aE_i^bL^cE_j^d$ vanish, where $C_{abcd}$ is the Weyl tensor. We introduce their linear perturbations $\Omega_{ij}[h]=\delta(C_{abcd}L^aE_i^bL^cE_j^d)$, where $\delta$ denotes the first-order variation induced by $h_{ab}$, including the variation of the null frame. The resulting components are gauge invariant and form a symmetric trace-free tensor on the three-dimensional transverse space~\cite{Durkee:2010xq,Durkee:2010qu}. In particular, the trace condition gives $\Omega_{+-}=-\Omega_{33}/2$, leaving five independent components $(\Omega_{++},\Omega_{--},\Omega_{+3},\Omega_{-3},\Omega_{33})$ that describe the five gravitational-wave polarizations. 

\emph{Killing--Yano symmetry and angular harmonics.---}The KY three-form $f$, dual to the principal two-form $\mathcal H$, defines a first-order symmetry operator acting on metric perturbations as $(\mathcal Q h)_{ab}= 2f_{(a}{}^{cd} \nabla_c h_{b)d}$~\cite{Frolov:2006dqt,Michishita:2019eeh}. On this Ricci-flat background, the operator $\mathcal Q$ constructed from $f$ maps each solution $h$ of the linearized Einstein equation to another solution $\mathcal Qh$~\cite{Michishita:2019eeh}. Evaluating $\Omega_{ij}[\mathcal Qh]$ and using $\delta R_{ab}[h]=0$, we find
\begin{equation} \label{eq:angular}
\Omega[\mathcal Qh]=F\Omega[h] \,.
\end{equation}
where $F$ is a first-order angular differential operator acting on the gauge-invariant Weyl components $\Omega_{ij}$.

The curvature identity~\eqref{eq:angular} is key to separating the perturbation equations. Suppressing the Fourier factor, we seek separated waves of the form
\begin{equation}
 \Omega_{ij}=R(r)A_{ij}(y).
\end{equation}
The five angular functions $A_{ij}(y)$, assembled into
$A(y)=(A_{++},A_{--},A_{+3},A_{-3},A_{33})^T$, describe the angular dependence of the gravitational-wave polarizations, while $R(r)$ describes their common radial dependence. Since $F$ involves only angular derivatives, we have $F\Omega=R(r)FA(y)$. To further distinguish the angular modes, we construct a second-order angular operator $J_1$ satisfying $[F,J_1]=0$. We then seek joint angular eigenfunctions satisfying
\begin{equation}\label{eq:joint}
 FA=\nu A\,,\qquad J_1A=\lambda A \,.
\end{equation}
The angular operators take the form $F=\boldsymbol{\alpha}_1\partial_y +\boldsymbol{\alpha}_0$ and $J_1=YI\partial_y^2+\boldsymbol{\beta}_1\partial_y +\boldsymbol{\beta}_0$, where $I$ is the $5\times5$ identity matrix. The eigenvalues $\lambda$ and $\nu$ are not independent but satisfy $H(\lambda,\nu)=0$. The explicit operator coefficients and the polynomial $H$ are given in the Supplemental Material (SM). The joint eigenvalue equations for $F$ and $J_1$, supplemented by smoothness at both rotation axes, determine the angular profiles of all five components $A_{ij}(y)$. Despite the different backgrounds and field equations, it is interesting to note that a similar organization of coupled polarizations by commuting angular operators appears in the separation of massive spin-two fields on Kerr~\cite{Dias:2026zri}.

We may also formulate the joint angular equations in terms of a single scalar, $s=2A_{++}/Y$. For $\omega\ne0$ and fixed $\lambda$, every root $\nu$ of $H(\lambda,\nu)=0$, including a repeated root, determines a two-dimensional space of angular solutions. At $\omega=0$, these roots are simple for the values of $\lambda$ admitted by smoothness at both rotation axes. Wherever $(s,s')$ uniquely parametrizes this solution space, all five angular components can be recovered as $A=u_0s+v_0s'$, and the joint equations~\eqref{eq:joint} reduce to a single second-order scalar equation,
\begin{equation}\label{eq:ang}
\frac{\dd^2s}{\dd y^2}=c_0(y)s+c_1(y)\frac{\dd s}{\dd y}\,.
\end{equation}
Here $u_0,v_0,c_0,c_1$ are determined algebraically from $F$ and $J_1$, as detailed in the SM. If this scalar parametrization fails at an interior latitude, the coupled angular equations~\eqref{eq:joint} remain regular and determine the continuation of $A$. The normalization $2/Y$ accounts for the frame transformation under the Carter interchange, allowing the scalar angular equation to be continued into the radial equation for $R(r)$.

\emph{Radial separation.---} The radial equation can be obtained from the angular equation through the Carter interchange
\begin{equation}\label{eq:swap}
\begin{split}
\widetilde r&=-\ii y \,,\quad \widetilde y=\ii r \,,\quad \widetilde\mu=-\mu \,,\\
\widetilde a^2+\widetilde b^2&=a^2+b^2-\mu \,,\quad \widetilde a\widetilde b=ab \,.
\end{split}
\end{equation}
This transformation exchanges the metric functions $X(r)$ and $Y(y)$. Keeping the Fourier labels $(\omega,k_P,k_w)$ fixed, the eigenvalues of $F$ and $J_1$ transform as $\widetilde\nu=\nu$ and $\widetilde\lambda=\lambda+\mu\omega^2$, respectively. Applying this continuation to~\eqref{eq:ang} yields
\begin{equation}\label{eq:rad}
\frac{\dd^2R}{\dd r^2}=-\widetilde c_0(\ii r)R+\ii\widetilde c_1(\ii r)\frac{\dd R}{\dd r}\,,
\end{equation}
where $\widetilde c_0$ and $\widetilde c_1$ are obtained from the angular coefficients with the continued parameters. Note that, the same radial equation can also be derived from the differential relations among the five curvature components implied by the linearized Einstein equations. Substituting $\Omega_{ij}=R(r)A_{ij}(y)$ and eliminating angular derivatives using~\eqref{eq:joint} reproduces~\eqref{eq:rad}.

At fixed $\nu\ne0$, the spectral relation $H(\lambda,\nu)=0$ is quadratic in $\lambda$. For generic parameters, its two distinct roots label two angular branches and, through the Carter interchange, two radial branches. Of the four candidate pairings, only the two satisfying $\widetilde\lambda=\lambda+\mu\omega^2$ obey the linearized vacuum Einstein equations on a regular reconstruction patch (see the SM).

\emph{Angular completeness.---}The angular eigenvalue equations must be supplemented by regularity conditions at the two rotation axes. At $y=a$ the $\phi$ circle collapses, whereas at $y=b$ the $\psi$ circle collapses. Since the complex transverse frame $E_i$ is not regular at these axes, smoothness is imposed on the curvature tensor represented by $\Omega_{ij}$, expressed in a smooth local Cartesian frame, rather than on its individual components in the Carter frame. For fixed $(\omega,m_\phi,m_\psi)$, these conditions select the allowed joint eigenvalues $(\lambda,\nu)$.

For real $\omega$, the angular operators $F$ and $J_1$ are symmetric with respect to the positive inner product
\begin{equation}\label{eq:angular-inner}
 \langle A,B\rangle_{\rm ang}
 =\int_b^a y\,\dd y\,A^\dagger\mathsf W B\,,
\end{equation}
where $A$ and $B$ are five-component angular functions with the same Fourier labels, and $\mathsf W=\operatorname{diag}(1,1,2,2,\tfrac32)$ is determined by the norm of the transverse curvature tensor $\Omega_{ij}$. As shown in the SM, the smooth-axis realization of $J_1$ is self-adjoint and elliptic on the compact angular three-sphere. Since $[F,J_1]=0$, the symmetric operator $F$ can be diagonalized within each finite-dimensional eigenspace of $J_1$. The resulting regular joint angular harmonics form a complete orthonormal basis.

Consequently, for real $\omega$, any smooth field with fixed Fourier labels admits the expansion
\begin{equation}\label{eq:harmonics}
\Omega(r,y)=\sum_j R_j(r)A^{(j)}(y) \,,
\end{equation}
where $\{A^{(j)}\}$ is a complete orthonormal basis of joint angular harmonics, including possible degeneracies, and $R_j(r)=\langle A^{(j)}, \Omega(r,\cdot) \rangle_{\rm ang}$. If $\Omega$ arises from a vacuum metric perturbation, its projection onto each joint $(\lambda,\nu)$ eigenspace also satisfies the differential constraints implied by the linearized Einstein equations on the regular reconstruction patches specified in the SM.

\emph{Radial boundary conditions.---} The future event horizon is generated by the Killing vector $\chi=(r_+^2\partial_T+\partial_P+ab\,r_+^{-2}\partial_w)/\mu$, with unit coefficient of the asymptotically normalized time translation $\partial_t$. Its action on a Fourier mode defines the frequency relative to the rotating horizon, $\omega_H=\omega-am_\phi/ (r_+^2+a^2)-bm_\psi/(r_+^2+b^2)$~\cite{Frolov:2002xf}. For a nonextremal horizon, the surface gravity is $\kappa_H= \Delta'(r_+)/ (2\mu r_+^2)$. For nonzero frequencies, the radial amplitude of a perturbation that is purely ingoing at the future horizon and purely outgoing at spatial infinity has the asymptotic behavior
\begin{equation}
R(r)\sim
\begin{cases}
(r-r_+)^{-\frac{ \ii\omega_H}{2\kappa_H} }, & r\to r_+\quad\text{(ingoing)},\\[2pt]
r^{-3/2}e^{\ii\omega r}, & r\to\infty\quad\text{(outgoing)}.
\end{cases}
\end{equation}
Together with the regularity conditions for the angular eigenfunctions at both rotation axes, these radial boundary conditions determine the complex quasinormal-mode frequencies. At $\omega=0$, regularity at the future horizon is imposed in ingoing coordinates, while the outgoing-wave condition at spatial infinity is replaced by the appropriate power-law behavior.

\emph{Recovering the metric perturbation.---}Although the five curvature components admit separated equations, not every separated solution necessarily corresponds to a vacuum metric perturbation. Additional compatibility conditions on $\Omega_{ij}$ are therefore required. As shown in the SM, a rank analysis yields ten independent fourth-order differential conditions that are locally sufficient for reconstructing a vacuum metric perturbation up to gauge. We choose the gauge $h_{TT}=h_{TP}=h_{Tw}=h_{PP}=h_{Pw}=0$, under which the curvature components and their derivatives up to fourth order determine the remaining ten metric components wherever the reconstruction matrices are nonsingular.

For a separated mode $\Omega_{ij}=R(r)A_{ij}(y)$, the higher derivatives can be eliminated using the scalar angular and radial equations~\eqref{eq:ang} and \eqref{eq:rad}. The gauge-fixed Fourier amplitude then takes the form
\begin{equation}\label{eq:h}
\mathfrak h_{ab}^{\rm gf}=\bm M_{ab}(r,y) \bigl (Rs,\,Rs',\, R's,\,R's' \bigr)^{\mathsf T}\,.
\end{equation}
Here $\bm M_{ab}$ is determined by the differential relations between the curvature components and the metric perturbation, together with the linearized Einstein equations, as detailed in the SM.

The compatibility conditions select the physical angular--radial pairings discussed above. On each physical branch, the two independent angular and radial solutions yield four local separated solutions. The reconstructed metric reproduces the prescribed curvature $\Omega_{ij}=RA_{ij}$ and satisfies the linearized vacuum Einstein equations $\delta R_{ab}[\mathfrak{h_{ab}}^{\rm gf}]=0$, confirming that these solutions describe genuine gravitational perturbations. For a fixed angular harmonic, the radial initial data $(R,R')$ at an ordinary point of~\eqref{eq:rad} determine the metric perturbation locally up to gauge. Perturbations of the black hole mass and angular momenta with $\Omega_{ij}=0$ require separate treatment.

\emph{Single-spin limit.---} The singly spinning limit provides both a nontrivial consistency check with Santos~\cite{Santos:2026pbu} and an illustration of the broader perturbation sectors covered by our construction. To make direct contact with Santos's results, we set $b=m_\psi=0$ and restrict to perturbations even under $\psi\mapsto-\psi$. Then $k_w=0$, $Y=a^2-y^2$, and the three even components are $(A_{--},A_{++},A_{33})$. Since $J_1$ preserves parity whereas $F$ reverses it, the angular operators acting within this sector are $(J_1)_{\rm e}$ and $(F^2)_e$. Introducing $x=y/a=\cos\theta$ and $D=\operatorname{diag}(1/2,1/2,1)$, we obtain the exact operator identities
\begin{equation}\label{eq:Santos-operators}
\begin{split}
D^{-1}(J_1)_{\rm e}D &= K_C-(m_\phi^2+3)I \,, \\
D^{-1}(F^2)_eD &= 3K_3-K_C+\big[(a\omega-m_\phi)^2+6\big]I \,.
\end{split}
\end{equation}
Here $K_C$ and $K_3$ are the angular operators of~\cite{Santos:2026pbu}. The corresponding separation constants are related by
\begin{equation}\label{eq:Santos-eigenvalues}
\begin{split}
\lambda_S &= \lambda+m_\phi^2+3 \,,\\
3\nu_S &= \nu^2+\lambda_S-(a\omega-m_\phi)^2-6 \,.
\end{split}
\end{equation}
With this identification and the Fourier-label reversal $(\omega_S,m_S)=(-\omega,-m_\phi)$ required by the different principal-null conventions, our angular spectral relation and second-order radial equation agree exactly with~\cite{Santos:2026pbu}.

The angular operators also have a singly spinning limit in the reflection-odd sector $(A_{+3},A_{-3})$ at $m_\psi=0$ and for nonzero $m_\psi$.

\emph{Angular symmetries in higher dimensions.---}We extend the angular construction to asymptotically flat vacuum MP black holes in $D=2n$ and $D=2n+1$ dimensions, with $n\geq2$ and nonzero spins of pairwise distinct magnitudes. We use canonical coordinates adapted to the principal tensor~\cite{Frolov:2006dqt,Krtous:2008tb} and a null frame aligned with the principal null directions. We work where these coordinates and the frame specified in the SM are regular in the exterior connected to spatial infinity. In odd dimensions we take the kernel of the principal tensor to be spacelike. Fix the frequency and azimuthal mode numbers, denoted collectively by $\boldsymbol{k}$. With $L$ the chosen principal null direction, the gauge-invariant curvature components $\Omega_{ij}[h]=\delta(C_{LiLj})$ form a symmetric trace-free tensor with $D(D-3)/2$ independent components. For any solution $h$ of the linearized vacuum Einstein equations in this region, the second-order symmetry operators associated with the Killing tensors~\cite{Krtous:2006qy}, denoted by $\mathcal S_j$, induce the angular operators $J_{j,\boldsymbol{k}}$ through
\begin{equation}
\begin{split} \label{eq:odd-commuting-main}
\Omega[\mathcal S_j h]&=J_{j,\boldsymbol{k}}\Omega[h] \,,\qquad
       \partial_r J_{j,\boldsymbol{k}}=0 \,,\\
[J_{i,\boldsymbol{k}},J_{j,\boldsymbol{k}}]&=0 \,,\qquad
       1\leq i,j\leq n-1 \,.
\end{split}
\end{equation}
The operators $J_{j,\boldsymbol{k}}$ differentiate only with respect to the latitude coordinates, and their coefficients are independent of $r$. In odd dimensions the KY three-form gives a first-order symmetry $\mathcal Q$ whose curvature action is
\begin{equation}
 \begin{split} \label{eq:hd-odd-ky-angular}
  \Omega[\mathcal Qh]&=F_{\boldsymbol{k}}\Omega[h],\qquad
       [F_{\boldsymbol{k}},J_{j,\boldsymbol{k}}]=0 \,,\\
  F_{\boldsymbol{k}}&=\sum_{\alpha=1}^{n-1}
    \boldsymbol{\alpha}_1^{(\alpha)}(\boldsymbol{x})\partial_{x_\alpha}
    +\boldsymbol{\alpha}_0(\boldsymbol{x}) \,.
 \end{split}
\end{equation}
Here $\boldsymbol{x}=(x_1,\ldots,x_{n-1})$ denotes the latitude coordinates. The first- and zeroth-order coefficients of $F_{\boldsymbol{k}}$ are matrices acting on the curvature components and, at fixed $\boldsymbol{k}$, depend only on $\boldsymbol{x}$. The SM constructs the angular operators and derives the curvature identities. In five dimensions, $J_{1,\boldsymbol{k}}$ is obtained from the operator $J_1$ in~\eqref{eq:joint} by reversing its sign and adding a scalar multiple of the identity (see the SM).

Table~\ref{tab:mp-angular-symmetries} lists the angular operators. In each dimension their number equals the number of independent rotation parameters. The $D=4$ member is Kerr, for which the separated curvature equations are those of Teukolsky~\cite{Teukolsky:1973ha}. 

Consider a mode satisfying $\mathcal S_j h=\lambda_j h+\mathcal L_{X_j}g$ for each $j$, where the Lie derivative is a pure gauge perturbation. Gauge invariance gives $J_{j,\boldsymbol{k}}\Omega[h]=\lambda_j\Omega[h]$ at every radius. In odd dimensions we may also select an eigenvalue $\nu$ of $F_{\boldsymbol{k}}$ by imposing $\mathcal Qh=\nu h+\mathcal L_X g$. Since the angular operators are independent of radius, we can expand the curvature at each radius in the same basis of angular eigenfunctions $A_{ij}^{(b)}(\boldsymbol{x})$ with these eigenvalues,
\begin{equation}
\Omega_{ij}(r,\boldsymbol{x}) =\sum_b R_b(r) A_{ij}^{(b)}(\boldsymbol{x}).\label{eq:hd-joint-angular-expansion}
\end{equation}
If the joint eigenvalues fix a unique angular profile up to normalization, all curvature components of the mode share one radial amplitude. If several profiles share these eigenvalues, the linearized Einstein equations determine whether their radial amplitudes evolve independently. We conjecture that the radial equations decouple in a suitable angular basis independent of $r$, so that each mode in this basis takes the form $\Omega_{ij}(r,\boldsymbol{x})=R(r)A_{ij}(\boldsymbol{x})$.
\begin{table}[tb]
\centering
\caption{Angular symmetries for gravitational perturbations of MP black holes. The arrows pair symmetry operators on the metric with the angular operators they induce on the curvature. The table uses the normalization of~\eqref{eq:odd-commuting-main}, with the Fourier labels suppressed. In the last two rows, $j=1,\ldots,n-1$.}
\label{tab:mp-angular-symmetries}
\setlength{\tabcolsep}{6pt}
\renewcommand{\arraystretch}{1.1}
\begin{tabular}{@{}cccc@{}}
\hline\hline
$D$ & Spins & First order & Second order\\
\hline
$4$ (Kerr) & $1$ & --- & $\mathcal S_1\to J_1$\\
$5$ & $2$ & $\mathcal Q\to F$ & $\mathcal S_1\to J_1$\\
$6$ & $2$ & --- & $\mathcal S_1\to J_1,\ \mathcal S_2\to J_2$\\
$7$ & $3$ & $\mathcal Q\to F$ & $\mathcal S_1\to J_1,\ \mathcal S_2\to J_2$\\
$\vdots$ & & & \\
$2n$ & $n-1$ & --- & $\mathcal S_j\to J_j$\\
$2n+1$ & $n$ & $\mathcal Q\to F$ & $\mathcal S_j\to J_j$\\
\hline\hline
\end{tabular}
\end{table}
%

\emph{Discussion.---}We have established the separation of gravitational perturbations of five-dimensional MP black holes with two nonzero spins of unequal magnitude. A key observation is the remarkably simple identity~\eqref{eq:angular}, which shows that the KY symmetry acts on the five gauge-invariant curvature components through a purely angular differential operator $F$. Together with a second, commuting operator $J_1$, this structure allows the coupled gravitational perturbation equations to be separated. The joint angular eigenvalue problem~\eqref{eq:joint} determines the relative profiles of the five curvature polarizations, while their radial dynamics reduce to a single second-order scalar equation. Thus, despite the coupling between gravitational polarizations, the perturbation equations remain separable, revealing an elegant underlying structure in the linearized Einstein equations and providing a practical framework for black hole spectroscopy. We have also discussed angular completeness at real frequencies, radial boundary conditions, and local metric reconstruction. In the singly spinning limit with $m_\psi=0$, the restrictions of $J_1$ and $F^2$ to the reflection-even sector provide a KY interpretation of Santos's angular operators in five dimensions~\cite{Santos:2026pbu}, as shown in~\eqref{eq:Santos-operators}.

Our construction further reveals a common angular symmetry structure for MP black holes in arbitrary dimensions with nonzero spins of pairwise distinct magnitudes. Second-order symmetry operators associated with the Killing tensors induce mutually commuting angular operators on the curvature components, with coefficients independent of radius. In odd dimensions, a KY three-form supplies an additional commuting first-order angular operator. This structure motivates the conjecture that the physical radial dynamics reduces to independent scalar equations in a suitable joint angular basis that is independent of radius. Establishing this conjecture requires resolving possible mixing within degenerate joint eigenspaces and deriving the corresponding radial equations. Complete separation also requires reducing the dependence on the individual latitude variables to ordinary differential equations, together with extending angular completeness and metric reconstruction beyond five dimensions. Six and seven dimensions provide natural settings in which to pursue these questions. Cases with coincident spin magnitudes require separate treatment because the canonical coordinates degenerate and the rotational symmetry is enhanced. In five dimensions the enhanced $U(2)$ symmetry allows an expansion in Wigner harmonics~\cite{Murata:2008yx,Tsuchiya:2020hur}.

The separated equations provide a framework for investigating gravitational quasinormal modes and stability across the two-spin parameter space of five-dimensional MP black holes, as well as for exploring non-Hermitian exceptional points, excitation factors, and resonant response~\cite{Motohashi:2024fwt,Zhao:2026nce}. Several extensions are worth exploring. One is to include a cosmological constant and examine whether the separation mechanism persists for MP black holes in (anti-)de Sitter spacetime~\cite{Chen:2006xh}. Another is to extend the construction to charged Chong--Cveti\v{c}--L\"u--Pope black holes, whose generalized KY symmetry with torsion suggests a possible route to separating gravitoelectromagnetic perturbations~\cite{Chong:2005hr,Kubiznak:2009qi}.

\par\medskip
\emph{Acknowledgments.---}We thank Li Li, Peng-Cheng Li, and Tieguang Zi for helpful discussions. J.-K.~Z. is particularly grateful to Li Li for his encouragement during this project. This work was supported in part by the National Natural Science Foundation of China (Grant Nos.~12505087, 12447134 and~12365011). H.-D.~L. was also supported by Postdoctoral Innovation Project of Shandong Province (SDCX-ZG-202503036).

\bibliography{refs}

\end{document}


\title{Supplemental Material for \textit{Separation of gravitational perturbations of Myers-Perry black holes}}
\author{Jun-Kun Zhao}
\email{junkunzhao@itp.ac.cn}
\affiliation{Institute of Theoretical Physics, Chinese Academy of Sciences, Beijing 100190, China}

\author{Hong-Da Lyu}
\email{hongdalyu@buaa.edu.cn}
\affiliation{Key Laboratory of Particle Physics and Particle Irradiation (MOE), \\Institute of Frontier and Interdisciplinary Science, \\Shandong University, Qingdao, Shandong 266237, China}

\maketitle
\section{Carter coordinates and curvature perturbations}
\paragraph{Frame and curvature conventions.}
The metric duals of the null frame vectors defined in the main text are
\begin{equation}\label{eq:dual-frame}
L^\flat=-\mathcal A-\frac{\Sigma}{X}\dd r \,,\qquad
N^\flat=-\frac{X}{2\Sigma}\left(\mathcal A-\frac{\Sigma}{X}\dd r\right) \,,\qquad
E_\pm^\flat=\frac{\sqrt{Y/2}}{p_\pm}\left(-\mathcal B\mp\ii\frac{\Sigma}{Y}\dd y\right) \,,\qquad
E_3^\flat=ry\,\mathcal C \,,
\end{equation}
where $p_\pm=r\pm\ii y$.

Setting $p_3=r$, the null frame satisfies $[L,E_i]=E_i/p_i$, while the only nonvanishing transverse component of $\dd L^\flat$, up to antisymmetry, is $\dd L^\flat(E_+,E_-)=2\ii y/\Sigma$. Since the transverse inner products are constant, the Koszul formula gives
\begin{equation}\label{eq:screen-transport}
2g(\nabla_L E_i,E_j) =\dd L^\flat(E_i,E_j) +\left(\frac{1}{p_i}-\frac{1}{p_j}\right)g(E_i,E_j)=0 \,.
\end{equation}
Thus $(\nabla_L E_i)_\perp=0$, where $\perp$ denotes projection onto the transverse space orthogonal to $L$ and $N$.

The curvature perturbation $\Omega_{ij}[h]=\delta[C(L,E_i,L,E_j)]$ includes the variation of the null frame. The perturbed null condition, $h_{LL}+2g(\delta L,L)=0$, fixes the part of $\delta L$ along $N$ to $\tfrac12 h_{LL}N$, where $h_{LL}=h_{ab}L^aL^b$. Since the background Weyl components of boost weights $+2$ and $+1$ vanish, the surviving frame contributions are obtained by replacing either occurrence of $L$ in the curvature projection with this term. For a vacuum perturbation, we therefore obtain
\begin{equation}
\Omega_{ij}[h]=\delta R(L,E_i,L,E_j)
+\frac12 h_{LL}\left[
R(N,E_i,L,E_j)+R(L,E_i,N,E_j)
\right].
\label{eq:curvature}
\end{equation}
Here $\delta R$ denotes the variation of the fully covariant Riemann tensor $R_{abcd}$, contracted with the fixed background frame. We use the convention $[\nabla_c,\nabla_d]v^a=R^a{}_{bcd}v^b$ and $R_{abcd}=g_{ae}R^e{}_{bcd}$.

\paragraph{Kaluza--Klein reduction.}
The kernel of the principal two-form $\mathcal H=-r\,\dd r\wedge\mathcal A+y\,\dd y\wedge\mathcal B$ is generated by $\partial_w$. Since $(\partial_w)^\flat=r^2y^2\mathcal C$, the four-dimensional Einstein-frame metric obtained by local Kaluza--Klein reduction is $ry(g_5-r^2y^2\mathcal C^2)$ up to an overall constant~\cite{Rasheed:1995zv}. The four-dimensional metric $g_5-r^2y^2\mathcal C^2$ has the Carter form of Petrov type D~\cite{Carter:1968ks,Krtous:2008tb}.

\section{Angular symmetry operators}
We choose the orientation $\epsilon_{TryPw}=ry\Sigma$, for which $\epsilon^{TryPw}=-1/(ry\Sigma)$. The Killing--Yano three-form defining the metric symmetry operator $\mathcal Q$ is~\cite{Frolov:2006dqt}
\begin{equation}
f=\star\left(
-r\,\dd r\wedge\mathcal A
+y\,\dd y\wedge\mathcal B
\right).
\label{eq:ky-form}
\end{equation}

\paragraph{Angular operators.}
The coefficients of $F$ follow from the curvature projection of $\mathcal Q h$. We construct $J_1$ with leading term $YI\partial_y^2$ and choose its remaining coefficients to satisfy $[F,J_1]=0$. Define the angular Fourier combination
\begin{equation}
W_y=-\omega y^2-k_P+\frac{abk_w}{y^2}
=Y\left(
\omega-\frac{am_\phi}{a^2-y^2}
+\frac{bm_\psi}{y^2-b^2}
\right).
\label{eq:W0}
\end{equation}
In the normalized null frame, with components ordered as $(++,--,+3,-3,33)$, the operators take the form
\begin{equation}
F=\boldsymbol\alpha_1\partial_y+\boldsymbol\alpha_0 \,,\qquad J_1=YI\partial_y^2+\boldsymbol\beta_1\partial_y+\boldsymbol\beta_0 \,,
\label{eq:unit-operators}
\end{equation}
where $I$ is the $5\times5$ identity matrix and primes on angular quantities denote derivatives with respect to $y$. The coefficient matrices can be expressed in terms of $\boldsymbol\alpha_1$ and the constant diagonal matrix $\boldsymbol{\Xi}=\operatorname{diag}(2,-2,1,-1,0)$. The matrix $\boldsymbol\alpha_1$ is
\begingroup\renewcommand{\arraystretch}{1.1}
\begin{equation}\label{eq:first}
\boldsymbol\alpha_1=\sqrt{\frac{Y}{2}}
\begin{pmatrix}
0 & 0 & 2 & 0 & 0 \\
0 & 0 & 0 & 2 & 0 \\
-1 & 0 & 0 & 0 & \frac{3}{2} \\
0 & -1 & 0 & 0 & \frac{3}{2} \\
0 & 0 & -2 & -2 & 0
\end{pmatrix}.
\end{equation}
\endgroup
The remaining coefficient matrices are
\begin{equation}\label{eq:unit-slots}
\boldsymbol\alpha_0=\frac{ab}{y^2}(6I-2\boldsymbol{\Xi}^{2})+\frac{k_w}{y}\boldsymbol{\Xi}+\frac{1}{2}\left(\frac{1}{y}+\frac{Y'}{2Y}\right)\boldsymbol\alpha_1 +\frac{1}{2}\left(\frac{1}{y}-\frac{Y'}{2Y}\right)[\boldsymbol{\Xi}^{2},\boldsymbol\alpha_1]+\frac{W_y}{Y}[\boldsymbol{\Xi},\boldsymbol\alpha_1] \,, 
\end{equation}
\begin{equation}\label{eq:beta1-matrix}
\boldsymbol\beta_1=\left(Y'+\frac{Y}{y}\right)I-\frac{2ab}{y^2}\boldsymbol\alpha_1 \,,
\end{equation}
and
\begin{equation}\label{eq:beta0-matrix}
\begin{aligned}
\boldsymbol\beta_0={}&\left(d_{33}-\frac{3Y}{y^2}\right)I+\left(G_0+\frac{Y}{2y^2}\right)\boldsymbol{\Xi}^{2}+G_1\boldsymbol{\Xi}-\frac{\boldsymbol\alpha_1^2}{y^2} +\frac{ab}{y^2}\left[\left(\frac{1}{y}-\frac{Y'}{2Y}\right)\boldsymbol\alpha_1+\left(\frac{1}{y}+\frac{Y'}{2Y}\right)[\boldsymbol{\Xi}^{2},\boldsymbol\alpha_1]\right]\\
&-\frac{2}{y^2}\left(\frac{abW_y}{Y}+k_w\right)[\boldsymbol{\Xi},\boldsymbol\alpha_1] \,.
\end{aligned}
\end{equation}
Here the brackets denote ordinary matrix commutators, and the scalar coefficients are
\begin{equation}
\begin{split} \label{eq:second}
d_{33}&=-\frac{W_y^2}{Y}-(a^2+b^2)\omega^2-2k_P\omega-\frac{k_w^2+3(a^2+b^2)}{y^2}-\frac{3a^2b^2}{y^4} \,, \\
G_0&=\frac{Y}{2y^2}-\frac{Y'^2}{4Y},\qquad G_1=\frac{W_yY'}{Y}+\frac{2abk_w}{y^3}+4\omega y \,.
\end{split}
\end{equation}

\paragraph{Curvature identity.}
To verify $\Omega[\mathcal Q h]=F\Omega[h]$, we evaluate both sides using the Riemann expression in~\eqref{eq:curvature} for arbitrary $h$ with fixed Fourier labels $(\omega,k_P,k_w)$. For vacuum perturbations, this expression agrees with the perturbed Weyl components $\Omega_{ij}$ defined in the main text~\cite{Durkee:2010qu}. Define $W_r=-\omega r^2+k_P+\frac{abk_w}{r^2}$ and the Ricci contractions $\delta R_{Li}=\delta R_{ab}[h]L^aE_i^b$ and $\delta R_{LL}=\delta R_{ab}[h]L^aL^b$. Writing $\mathcal R_{ij}=(F\Omega[h]-\Omega[\mathcal Q h])_{ij}$, direct calculation gives
\begin{equation}
\begin{split}
\mathcal R_{\pm\pm}&=0 \,,\\
\mathcal R_{\pm3}&=\frac{1}{y}\sqrt{\frac{Y}{2}}\bigg[\left(\frac{1}{2}-\frac{y^2}{p_\pm^2}\right)\delta R_{LL}+\frac{yp_\mp}{2p_\pm}\left(\partial_y\pm\frac{W_y}{Y}\right)\delta R_{LL}\bigg] +y\left(\partial_r+\frac{2}{p_\pm}-\frac{\ii W_r}{X}\right)\delta R_{L\pm} \,, \\
\mathcal R_{33}&=\left[2ab\left(\frac{1}{y^2}-\frac{1}{r^2}\right)+\frac{2\ii k_w}{r}\right]\delta R_{LL}+2y\left(\partial_r+\frac{2}{r}-\frac{\ii W_r}{X}\right)\delta R_{L3} \,.
\end{split}
\end{equation}
The derivatives act on $\delta R_{LL}$ and $\delta R_{Li}$, including the coordinate dependence of the frame vectors. For a vacuum perturbation, $\delta R_{ab}[h]=0$, so $\mathcal R_{ij}=0$ and hence $\Omega[\mathcal Q h]=F\Omega[h]$.

\paragraph{Spectral relation.}
Direct substitution of the coefficient matrices gives $[F,J_1]=0$ and $H(J_1,F)=0$, where
\begin{equation}
\begin{split} \label{eq:curve}
H(\lambda,\nu)&=\nu^5+(5\ell-3)\nu^3-42\omega k_w\nu^2+(4\ell^2-12\ell+72k_P\omega)\nu -24\omega k_w\ell+144ab\omega^2 \,,\\
\ell&=\lambda+(a^2+b^2)\omega^2+2k_P\omega \,.
\end{split}
\end{equation}
Consequently, the joint eigenvalues defined by $FA=\nu A$ and $J_1A=\lambda A$ satisfy $H(\lambda,\nu)=0$.

\paragraph{Spin reversals.}
Under either $(a,\phi,m_\phi)\to(-a,-\phi,-m_\phi)$ or $(b,\psi,m_\psi)\to(-b,-\psi,-m_\psi)$, the coordinate and Fourier relations leave $(T,P,r,y,\omega,k_P)$ unchanged and send $(w,k_w)\to(-w,-k_w)$. This transformation reverses $E_3$ and, with our orientation convention, changes the sign of the Killing--Yano three-form $f$. The angular operators therefore transform as
\begin{equation}\label{eq:reflection}
F\to-\boldsymbol{\Theta}F\boldsymbol{\Theta},\qquad J_1\to\boldsymbol{\Theta}J_1\boldsymbol{\Theta},
\end{equation}
where $\boldsymbol{\Theta}=\operatorname{diag}(1,1,-1,-1,1)$. The joint eigenvalues accordingly transform as $(\lambda,\nu)\to(\lambda,-\nu)$.

\section{Obtaining the angular and radial equations}
We derive the scalar angular equation from the joint eigenvalue problem and obtain the radial equation through the Carter interchange.
\paragraph{Scalar angular equation.}
Away from the rotation axes, where $Y\ne0$, the equation $J_1A=\lambda A$ gives
\begin{equation}\label{eq:angular-U}
A''=U_1A+U_2A' \,,\qquad U_1=\frac{\lambda I-\boldsymbol\beta_0}{Y} \,,\qquad U_2=-\frac{\boldsymbol\beta_1}{Y} \,.
\end{equation}
The equations for $A$ and $A'$ can be written as $\partial_y(A,A')^{\mathsf T}=D_y(A,A')^{\mathsf T}$. Similarly, the action of $F$ is represented by $(FA,(FA)')^{\mathsf T}=F_{\rm data}(A,A')^{\mathsf T}$, where
\begin{equation}\label{eq:angular-data}
D_y=\begin{pmatrix}0&I\\ U_1&U_2\end{pmatrix},\qquad
F_{\rm data}=\begin{pmatrix}
\boldsymbol\alpha_0 & \boldsymbol\alpha_1\\
\boldsymbol\alpha_0'+\boldsymbol\alpha_1U_1 & \boldsymbol\alpha_0+\boldsymbol\alpha_1'+\boldsymbol\alpha_1U_2
\end{pmatrix}.
\end{equation}
Since $[F,J_1]=0$, $FA$ also satisfies $J_1(FA)=\lambda FA$. It follows that
\begin{equation}\label{eq:data-commutator}
F_{\rm data}'=[D_y,F_{\rm data}],
\end{equation}
so the characteristic polynomial of $F_{\rm data}$ is independent of $y$. Substituting the explicit coefficient matrices gives
\begin{equation}\label{eq:characteristic}
\det(\nu I_{10}-F_{\rm data})=H(\lambda,\nu)^2,
\end{equation}
where $I_{10}$ is the $10\times10$ identity matrix. On solutions of $J_1A=\lambda A$, the operator identity $H(J_1,F)=0$ also gives $H(\lambda,F_{\rm data})=0$.

The condition $FA=\nu A$ requires the initial values to satisfy $F_{\rm data}(A,A')^{\mathsf T}=\nu(A,A')^{\mathsf T}$. At fixed $\lambda$, the two polynomial identities imply that each simple root of $H(\lambda,\nu)$ in $\nu$ gives two linearly independent solutions of the joint angular equations. This count is made before imposing smoothness at the rotation axes. For $\omega\ne0$, each repeated root also gives exactly two linearly independent solutions of the joint angular equations, as shown below.

We introduce $s=2A_{++}/Y$ and the vector $e_{++}=(1,0,0,0,0)^{\mathsf T}$, so that $e_{++}^{T}A=A_{++}$. The scalar variable and its derivative satisfy $(s,s')^{\mathsf T}=U_0(A,A')^{\mathsf T}$, where
\begingroup\renewcommand{\arraystretch}{1.2}
\begin{equation}\label{eq:normalized-basis}
U_0=\frac{2}{Y}\begin{pmatrix}
e_{++}^{\mathsf T}&0\\
-\frac{Y'}{Y}e_{++}^{\mathsf T}&e_{++}^{\mathsf T}
\end{pmatrix} \,.
\end{equation}
\endgroup
On a branch with two independent angular solutions, the $10\times2$ reconstruction matrix $L_0$ is determined by
\begin{equation}\label{eq:scalar-inverse}
(F_{\rm data}-\nu I_{10})L_0=0 \,,\qquad U_0L_0=I_2 \,,
\end{equation}
where $I_2$ is the $2\times2$ identity matrix. These equations uniquely determine $L_0$ wherever two independent angular solutions yield linearly independent pairs $(s,s')$ at that latitude. An explicit expression for $L_0$ follows from the spectral polynomial. For $H(\lambda,\nu)=0$, define
\begin{equation}
\begin{split}
\boldsymbol{\Phi}_\nu={}&F_{\rm data}^{4}+\nu F_{\rm data}^{3}+(\nu^{2}+5\ell-3)F_{\rm data}^{2} +\bigl[\nu^{3}+(5\ell-3)\nu-42\omega k_w\bigr]F_{\rm data} \\
&+\bigl[\nu^{4}+(5\ell-3)\nu^{2}-42\omega k_w\nu+4\ell^{2}-12\ell+72k_P\omega\bigr]I_{10} \,.
\end{split}
\end{equation}
This is the matrix polynomial obtained by dividing $H(\lambda,z)$ by $z-\nu$ and then substituting $z=F_{\rm data}$. It satisfies $(F_{\rm data}-\nu I_{10})\boldsymbol{\Phi}_\nu=0$. For a root of multiplicity $m$, the two Jordan blocks associated with $\nu$ both have size $m$, as established below. The polynomial therefore has rank two and image $\ker(F_{\rm data}-\nu I_{10})$. Write $\Phi_{ij}=(\boldsymbol{\Phi}_\nu)_{ij}$ and choose two columns $p,q\in\{1,\ldots,10\}$ such that $\Delta_\nu=\Phi_{1p}\Phi_{6q}-\Phi_{1q}\Phi_{6p}\ne0$. The entries of $L_0$ are then
\begin{equation}
(L_0)_{i1} =\frac{(Y\Phi_{6q}-Y'\Phi_{1q})\Phi_{ip}-(Y\Phi_{6p}-Y'\Phi_{1p})\Phi_{iq}}{2\Delta_\nu} \,,\qquad
(L_0)_{i2} =\frac{Y(\Phi_{1p}\Phi_{iq}-\Phi_{1q}\Phi_{ip})}{2\Delta_\nu} \,,
\qquad i=1,\ldots,10  \,.
\end{equation}

In that case, $(A,A')^{\mathsf T}=L_0(s,s')^{\mathsf T}$, and the scalar equation and angular reconstruction take the form
\begin{equation}\label{eq:scalar-angular}
\frac{\dd}{\dd y}\begin{pmatrix}s\\s'\end{pmatrix}
=\begin{pmatrix}0&1\\c_0&c_1\end{pmatrix}\begin{pmatrix}s\\s'\end{pmatrix} \,,\qquad A=u_0s+v_0s' \,.
\end{equation}
The coefficients are given by
\begin{equation}\label{eq:scalar}
\begin{pmatrix}0&1\\c_0&c_1\end{pmatrix}
=(U_0'+U_0D_y)L_0 \,,\qquad
(u_0,v_0)=(I,0)L_0 \,.
\end{equation}
The first identity follows by differentiating $(s,s')^{\mathsf T}=U_0(A,A')^{\mathsf T}$ and using~\eqref{eq:angular-data}; its second row gives $s''=c_0s+c_1s'$. Conversely, Equations~\eqref{eq:data-commutator} and~\eqref{eq:scalar-inverse} ensure that every solution of the scalar equation reconstructs a solution of both joint angular equations wherever $L_0$ is defined. At latitudes $b<y<a$ where $A$ and $A'$ cannot be recovered uniquely from $s$ and $s'$, the coupled angular equations remain regular and allow the angular solution $A$ to be continued smoothly across these points.

\paragraph{Repeated roots.}
For $\omega\ne0$, let $\nu$ be a root of $H(\lambda,\cdot)$ of multiplicity $m$. Equation~\eqref{eq:characteristic} gives algebraic multiplicity $2m$, while $H(\lambda,F_{\rm data})=0$ bounds the size of each Jordan block associated with $\nu$ by $m$, so $\dim\ker(F_{\rm data}-\nu I_{10})\ge2$. For the reverse bound, regard $k_P,k_w$ as independent local parameters and vary $\tau=ab$ at fixed $a^2+b^2,y,\omega,k_P,k_w,\lambda,\nu$. The matrix $F_{\rm data}$ is analytic in $\tau$ at an interior latitude, and $\partial_\tau H=144\omega^2\ne0$ implies that $\det(\nu I_{10}-F_{\rm data})=H^2$ has a zero of order exactly two at the parameter value under consideration. Since this order bounds the kernel dimension from above, the kernel is exactly two-dimensional.

\paragraph{Carter interchange and radial equation.}
Under the Carter interchange specified in the main text, the coordinates $(T,P,w)$ and Fourier labels $(\omega,k_P,k_w)$ remain fixed, while the background quantities transform as
\begin{equation}
\begin{split}
\widetilde X(\widetilde r)&=Y(y) \,,\qquad \widetilde Y(\widetilde y)=X(r) \,,\qquad \widetilde\Sigma=-\Sigma  \,,\\
\widetilde{\mathcal A}&=\mathcal B \,,\qquad \widetilde{\mathcal B}=\mathcal A \,,\qquad \widetilde{\mathcal C}=\mathcal C \,.
\end{split}
\end{equation}
These relations preserve the metric and the principal two-form. The coordinate Jacobian and the ratio $\widetilde r\widetilde y\widetilde\Sigma/(ry\Sigma)$ are both $-1$, so the oriented volume form is also preserved. Hence the KY three-form $f$ and the symmetry operator $\mathcal Q$ are unchanged.

Using $\widetilde p_+=-p_+$, the relevant frame covectors transform as
\begin{equation}
\widetilde L^\flat=p_+\sqrt{\frac{2}{Y}}E_+^\flat \,,\qquad \widetilde E_+^\flat=-\frac{1}{p_+}\sqrt{\frac{X}{2}}L^\flat \,.
\end{equation}
For this background, $R(N,E_+,L,E_+)=0$, so the explicit $h_{LL}$ term in~\eqref{eq:curvature} vanishes for $\Omega_{++}$. Using the transformations of $L^\flat$ and $E_+^\flat$ above, we then obtain
\begin{equation}\label{eq:interchange-curvature}
\widetilde\Omega_{++}=\frac{X}{Y}\Omega_{++} \,,\qquad \frac{2\widetilde\Omega_{++}}{X}=\frac{2\Omega_{++}}{Y} \,.
\end{equation}
Thus the normalized curvature component $2\Omega_{++}/Y=R(r)s(y)$ is invariant under the interchange, allowing the scalar angular equation to be continued into the radial equation.

We continue the angular equations with $\widetilde\nu=\nu$ and $\widetilde\lambda=\lambda+\mu\omega^2$, preserving $\ell$ and hence the spectral relation~\eqref{eq:curve}. Let $\widetilde s$ solve the scalar angular equation with the continued parameters and eigenvalues. Setting $R(r)=\widetilde s(\ii r)$ gives $R'(r)=\ii\widetilde s'(\ii r)$ and $R''(r)=-\widetilde s''(\ii r)$. Equation~\eqref{eq:scalar-angular} therefore yields
\begin{equation}\label{eq:radial}
R''(r)=-\widetilde c_0(\ii r)R(r)+\ii\,\widetilde c_1(\ii r)R'(r).
\end{equation}

At fixed $\nu\ne0$, the spectral relation~\eqref{eq:curve} is quadratic in $\lambda$ and generically has two distinct roots. The scalar equations alone do not determine which angular and radial branches can be combined into a vacuum metric perturbation. The reconstruction analysis below establishes the physical pairing $\widetilde\lambda=\lambda+\mu\omega^2$ by imposing the full linearized Einstein equations.

\section{Local metric reconstruction}
Fix the Fourier labels and work on an ordinary patch where the minors used below are nonzero. Let $h_N$ collect the fifteen metric components and their $(r,y)$-derivatives through total order $N$, and let $E_Nh_N$ collect the linearized Ricci tensor and its derivatives through order $N-2$. The map $G_N$ takes a gauge vector and its derivatives through order $N+1$ to metric data for a pure gauge perturbation, with $E_NG_N=0$. We write $\mathcal T_n h_{n+2}=j^n(T_p\Omega[h])$, where $j^n$ includes all derivatives of total order at most $n$ and $T_p=\mathrm{diag}(2p_+^2/Y,2p_-^2/Y,ryp_+/\sqrt{Y/2},ryp_-/\sqrt{Y/2},r^2y^2)$. The product rule makes this weighting invertible at every jet order. Set $t_N=(N+1)(N+2)/2$, $d_N=\dim(\ker E_N/\mathrm{im}\,G_N)$ and $s_N=\mathrm{rank}(\mathcal T_{N-2}|_{\ker E_N})$. Table~\ref{tab:jet-rank} gives the ranks used below.

\begin{table}[ht]
\caption{Ranks of the Ricci, gauge, and curvature maps at a generic ordinary point.}
\label{tab:jet-rank}
\centering
\begin{tabular}{rrrrr}
\toprule
$N$ & $\mathrm{rank}\,E_N$ & $\mathrm{rank}\,G_N$ & $d_N$ & $s_N$ \\
\midrule
4 & 75  & 105 & 45 & 30 \\
5 & 120 & 140 & 55 & 50 \\
6 & 175 & 180 & 65 & 65 \\
7 & 240 & 225 & 75 & 75 \\
8 & 315 & 275 & 85 & 85 \\
\bottomrule
\end{tabular}
\end{table}

The Bianchi identities give $\mathrm{rank}\,E_N\le 15t_{N-2}-5t_{N-3}$, while $\mathrm{rank}\,G_N\le5t_{N+1}$ and $s_N\le\min(5t_{N-2},d_N)$. These bounds are saturated by exact minors at $(a,b,\mu,\omega,m_\phi,m_\psi;r,y)=(2/3,1/2,2,3/5,2,0;3,3/5)$, reduced modulo $65537$ with $\ii=256$. The corresponding ranks hold wherever these minors remain nonzero.

Let $A_6$ be the matrix obtained by stacking $\mathcal T_4$ below $E_6$. Its rank is $175+65=240$. Since the metric six-jet has dimension $420$ and $A_6G_6=0$, its kernel is precisely the $180$-dimensional gauge image. Choose $175$ independent Ricci rows and $65$ curvature rows with a common invertible minor $M=A_6[I,L_c]$, selecting curvature rows in order of increasing derivative order. Let $b_I[\Omega]$ contain zeros in the Ricci slots and the prescribed curvature data in the remaining slots. The ten unused curvature rows $u$ give the necessary conditions
\begin{equation}
(\mathcal D\Omega)_u=[j^4(T_p\Omega)]_u-A_6[u,L_c]M^{-1}b_I[\Omega]=0 \,.
\label{eq:quartic}
\end{equation}
Fix $U(h)=(h_{TT},h_{TP},h_{Tw},h_{PP},h_{Pw})^{\mathsf T}=0$. On this Fourier mode, $U(\mathcal L_\xi g)=\mathsf D\xi$ is algebraic, with $\mathsf D_{AB,c}=\partial_c g_{AB}+\ii k_A g_{cB}+\ii k_B g_{Ac}$ and $k_T=-\omega$. Where $\mathsf D$ is invertible, let $V_h$ map a metric jet to the gauge-fixed components $h-\mathcal L_{\mathsf D^{-1}U(h)}g$. As $V_h$ annihilates $\ker A_6$, the gauge-fixed metric is
\begin{equation}
h^{\rm gf}=\mathcal R\Omega:=V_h|_{L_c}M^{-1}b_I[\Omega] \,.
\label{eq:inverse}
\end{equation}
Here $V_h|_{L_c}$ denotes the columns of $V_h$ selected by $L_c$. The reconstruction holds where the gauge matrix is invertible and the reconstruction minors are nonzero. The gauge matrix is singular when all three Fourier labels vanish. Stationary variations of the Myers--Perry parameters lie in the kernel of the curvature map.

To prove sufficiency, let $\Omega$ be $C^6$ and satisfy $\mathcal D\Omega=0$ throughout the patch. At orders four, five and six, exact minors give $10,20,25$ independent leading rows of the constraints and their derivatives. Their kernel in the curvature six-jets therefore has dimension at most $140-55=85$. It contains the curvature data generated by formal vacuum metric eight-jets, whose image already has dimension $s_8=85$. Thus, at each point there is a formal metric eight-jet $h_8$ satisfying $E_8h_8=0$ and $\mathcal T_6h_8=j^6(T_p\Omega)$. The identity $V_hh_6=V_h|_{L_c}M^{-1}(A_6h_6)_I$ holds on arbitrary metric six-jets. Differentiating it twice, including its coefficients, identifies the two-jet of $\mathcal R\Omega$ with that of the matching gauge-fixed metric jet. Consequently
\begin{equation}
\delta R_{ab}[\mathcal R\Omega]=0,\qquad \Omega[\mathcal R\Omega]=\Omega \,.
\end{equation}
For a compatible product $\Omega_{ij}=RA_{ij}$, write $2\Omega_{++}/Y=Rs$ on a regular scalar patch. The two scalar equations reduce all derivatives in $\mathcal R$ to the four quantities used in the main text,
\begin{equation*}
 \mathfrak h_{ab}^{\rm gf}=\bm M_{ab}(r,y)  (Rs,Rs',R's,R's')^{\mathsf T} \,.
\end{equation*}
The full gauge-fixed perturbation is $h_{ab}^{\rm gf}=e^{-\ii\omega T+\ii k_P P+\ii k_w w}\mathfrak h_{ab}^{\rm gf}$. For a fixed angular harmonic, $R$ and $R'$ determine this perturbation on the reconstruction patch.

\section{Physical angular--radial pairing}
Fix $\nu\ne0$ on a regular sheet where $H(\lambda,\nu)=0$ has distinct roots $\lambda_1,\lambda_2$. Let $\mathcal E_{ij}$ be the four-dimensional space spanned by products of the two radial solutions at $\widetilde\lambda_i=\lambda_i+\mu\omega^2$ and the two angular solutions at $\lambda_j$. For a vacuum $\mathcal Q$ eigenmode modulo gauge, the curvature identity gives $F\Omega=\nu\Omega$; the rank-four coefficient of $\partial_y$ and the remaining algebraic equation leave four angular initial data. The two $J_1$ branches exhaust these data. The Carter interchange gives four radial data for the same component $2\Omega_{++}/Y$. Both value-and-derivative maps through third order have rank four at the reference point below. Wherever they remain invertible, the two univariate fourth-order equations place every such curvature field in $\bigoplus_{ij}\mathcal E_{ij}$.

At the reference parameters above take $\nu=8/3$ and $(\lambda_1,\lambda_2)=(-3943/900,-563/180)$. The ranks of $\mathcal D$ on $(\mathcal E_{11},\mathcal E_{12},\mathcal E_{21},\mathcal E_{22})$ are $(0,4,4,0)$ at the reference point. The combined crossed block has rank eight. Both matched reconstructions satisfy the Ricci and curvature equations identically in $(r,y)$ at these parameter values, giving eight independent physical solutions on the reference patch.

To extend the count, choose a noncharacteristic slice $r=r_*$ and impose the five cyclic gauge conditions. The hypersurface constraints are algebraic in the momenta conjugate to these cyclic metric components. The canonical pairing between constraints and gauge transformations identifies their coefficient matrix with the transpose of $\mathsf D$ at opposite Fourier momentum, up to invertible lapse and shift factors. On the remaining ten functions of $y$, the physical action of $\mathcal Q-\nu$ is a square ordinary differential system with an eight-dimensional kernel at the reference parameters. Here vacuum curvature determines the metric modulo gauge and intertwines $\mathcal Q$ with $F$, so the reference matched products exhaust this kernel. Locally in $y$, row and column operations with differential operators with meromorphic coefficients give a diagonal system. The finite kernel rules out any zero diagonal operator, which would leave a free function. Each scalar equation can be solved for an arbitrary analytic source at an ordinary latitude. Successive differentiation in the parameters therefore lifts all eight solutions formally, using vacuum evolution from the slice at each order.

Evaluate the compatibility and reconstructed field-equation residuals, with their derivatives along both product evolutions, on the sixteen product data. The independent formal solutions force all $9\times9$ minors to have zero Taylor series. These are rational identities on the regular spectral sheet. The nonzero crossed $8\times8$ minor gives the opposite bound, so the common kernel is the graph of a rational map from $\mathcal E_{11}\oplus\mathcal E_{22}$ to $\mathcal E_{12}\oplus\mathcal E_{21}$. The kernel is preserved by both evolutions. Every graph block would thus give a rational differential map between different branches in at least one variable.

To exclude such maps, specialize the coefficient family at $\omega=0$ with generic $\nu$. With $\boldsymbol\alpha_{1,\infty}=\lim_{y\to\infty}\boldsymbol\alpha_1/y$, the leading Euler terms are $\boldsymbol\alpha_{1,\infty}(y\partial_y+1)$ and $-(y\partial_y)(y\partial_y+2)I+3I-\boldsymbol\alpha_{1,\infty}^2$. The algebraic row eliminates the zero eigendirection of $\boldsymbol\alpha_{1,\infty}$. For $A\sim y^\rho v$, the eigenvalues $\pm1$ and $\pm2$ give $(\lambda,\rho)=(3-\nu^2,-1\pm\nu)$ and $(-\nu^2/4,-1\pm\nu/2)$, respectively. All four leading eigenvectors have nonzero $++$ components, so the scalar readout shifts every power by $-2$; the Carter substitution preserves these powers in the radial equation. The two sets are disjoint modulo integers for generic $\nu$. A rational differential map preserves exponent classes modulo integers and hence vanishes between these branches; the frame factor $\sqrt{Y/2}$ is also meromorphic at infinity. The distinct roots and regular rank-two projectors allow branch frames regular at $\omega=0$. The first nonzero Laurent coefficient of a hypothetical generic map would then give a forbidden map between the specialized systems. All graph blocks therefore vanish, giving
\begin{equation}
\ker\mathcal D\big|_{\oplus_{ij}\mathcal E_{ij}}=\mathcal E_{11}\oplus\mathcal E_{22}
\label{eq:physical-pairing}
\end{equation}
on the generic sheet and by continuation wherever the required branches, projections and reconstruction ranks remain regular. For each fixed angular solution the two matched radial solutions have nonzero Wronskian on an ordinary scalar patch. Any monic second-order equation obtained there by direct elimination must consequently have the same coefficients.

For $\nu=0$ and $\omega k_w\ne0$, the spectral equation fixes $\ell=6ab\omega/k_w$. Since $\partial_\lambda H(\lambda,0)=-24\omega k_w\ne0$, the finite root is analytic in $\nu$. The angular and continued radial data have two-dimensional kernels at the limit by the nonzero-frequency rank result. A nonzero $8\times8$ minor in each ten-dimensional data system supplies analytic kernel bases. Wherever their scalar readouts and the metric reconstruction stay regular, the resulting four product solutions retain independent initial data. The matched Ricci and curvature identities extend to the finite branch by continuity. At $\omega=0$, $H(\lambda,0)=0$ for every $\lambda$, so the stationary $\nu=0$ sector requires a separate radial analysis.

\section{Angular completeness and physical spectral projections}
\paragraph{Angular operators and regularity.}
We fix real $\omega$ and integer azimuthal numbers $(m_\phi,m_\psi)$. The normalized transverse tensor determines the positive angular inner product
\begin{equation}
\langle A,B\rangle_{\rm ang}
=\int_b^a y\,\dd y\,A^\dagger\mathsf W B,
\qquad
\mathsf W=\operatorname{diag}(1,1,2,2,\tfrac32).
\label{eq:angular-weight}
\end{equation}
A geometric representation of $F$ and $J_1$ makes their regularity and spectral properties explicit. Introduce the auxiliary unit round metric
\begin{equation}
q=\frac{\dd y^2}{Y}
+\frac{a^2-y^2}{a^2-b^2}\,\dd\phi^2
+\frac{y^2-b^2}{a^2-b^2}\,\dd\psi^2.
\label{eq:round-angular-metric}
\end{equation}
Its volume measure is proportional to $y\,|\dd y|\,\dd\phi\,\dd\psi$. For $\mathcal K=\boldsymbol\alpha_1/\sqrt Y$, define
\begin{equation}
\mathsf G=e^{\eta\mathcal K},
\qquad
\cos\eta=\frac{a\sqrt{y^2-b^2}}{y\sqrt{a^2-b^2}},
\qquad
\sin\eta=\frac{b\sqrt{a^2-y^2}}{y\sqrt{a^2-b^2}}.
\label{eq:angular-frame-rotation}
\end{equation}
Since $\mathcal K^\dagger\mathsf W+\mathsf W\mathcal K=0$, this rotation satisfies $\mathsf G^\dagger\mathsf W\mathsf G=\mathsf W$.

The components $\mathsf G A$ represent a complex symmetric trace-free tensor $T_{ij}$ on $(S^3,q)$. We use the orthonormal frame $e_1=-\sqrt Y\,\partial_y$, $e_2=\sqrt{(a^2-b^2)/(a^2-y^2)}\,\partial_\phi$, and $e_3=\sqrt{(a^2-b^2)/(y^2-b^2)}\,\partial_\psi$, with complex vectors $e_\pm=(e_2\pm\ii e_1)/\sqrt2$. Denote the connection of $q$ by $\bar\nabla$ and choose $\bar\epsilon(e_1,e_2,e_3)=1$. Define the symmetric tensor curl and the multiplication operators $\mathsf M_{u,v}$ by
\begin{equation}
(\operatorname{curl}_2T)_{ij} =\bar\epsilon_i{}^{kl}\bar\nabla_kT_{lj}
+\bar\epsilon_j{}^{kl}\bar\nabla_kT_{li} \,,\qquad
(\mathsf M_{u,v}T)_{ij} =-\ii\left[
(\bar\nabla_iZ_{u,v}^{k})T_{kj}
+(\bar\nabla_jZ_{u,v}^{k})T_{ik}
\right] \,, \qquad Z_{u,v}=u\,\partial_\phi+v\,\partial_\psi \,.
\label{eq:round-tensor-operators}
\end{equation}
For real $u,v$, the Killing property of $Z_{u,v}$ makes $\mathsf M_{u,v}$ a smooth Hermitian endomorphism of the symmetric trace-free tensor bundle. No divergence-free condition is imposed on $T_{ij}$.

After restoring the azimuthal derivatives, direct substitution of the coefficient matrices gives
\begin{equation}\label{eq:geometric-angular-operators}
\mathsf G F(\omega)\mathsf G^{-1} =\operatorname{curl}_2-\omega\mathsf M_{b,a} \,,\qquad
\mathsf G J_1(\omega)\mathsf G^{-1} =\bar\nabla^k\bar\nabla_k-3I +2\omega\mathsf M_{a,b} +\omega^2y^2I \,.
\end{equation}
Here the frequency dependence is evaluated at fixed $(m_\phi,m_\psi)$, using the Fourier-label relations in the main text. Equation~\eqref{eq:geometric-angular-operators} identifies the angular operators with globally defined tensor operators on the round three-sphere, supplemented by smooth Hermitian potentials.

For completeness, the corresponding axis conditions can be stated directly in the original frame. Introduce
\begin{equation}
(\rho^2,\chi)=
\begin{cases}
\left(\dfrac{a^2-y^2}{a^2-b^2},\phi\right),
& y\to a,\\[6pt]
\left(\dfrac{y^2-b^2}{a^2-b^2},\psi\right),
& y\to b.
\end{cases}
\label{eq:axis-coordinates}
\end{equation}
The vectors $e^{\mp\ii\chi}E_\pm$ and $E_3$ are smooth in local Cartesian coordinates. Consequently, a component with azimuthal number $m$ along the collapsing circle is regular precisely when
\begin{equation}
A_j=\rho^{|m-\boldsymbol\Xi_{jj}|}b_j(\rho^2),
\qquad b_j\in C^\infty.
\label{eq:axis-regularity}
\end{equation}
The frame rotation~\eqref{eq:angular-frame-rotation} extends across both axes under these regular frame transitions, so the geometric representation preserves this smooth tensor domain.

The tensor Laplacian $\bar\nabla^k\bar\nabla_k$ is nonpositive, elliptic and self-adjoint on the compact three-sphere. Its smooth Hermitian perturbation in~\eqref{eq:geometric-angular-operators} therefore gives $J_1$ a self-adjoint realization with compact resolvent and finite-dimensional eigenspaces. The tensor curl is formally symmetric, so $F$ is also symmetric and preserves smoothness. Since $[F,J_1]=0$, its restriction to each $J_1$ eigenspace is a finite-dimensional Hermitian operator. Diagonalizing these restrictions gives a complete orthonormal basis of regular joint eigenfunctions in each integer Fourier sector.

At $\omega=0$, the tensor corresponding to an eigenfunction of $J_1$ satisfies
\begin{equation}
(-\lambda-3)\|T\|_q^2
=\|\bar\nabla T\|_q^2\ge0.
\label{eq:stationary-angular-bound}
\end{equation}
Thus $\lambda\le-3$, and all roots in $\nu$ of $H(\lambda,\nu)=\nu(\nu^2+4\lambda)(\nu^2+\lambda-3)$ are simple on the regular stationary spectrum.

\paragraph{Preservation of the vacuum constraints.}
Angular completeness must be supplemented by preservation of the compatibility conditions~\eqref{eq:quartic}. Let $\mathcal D$ denote these ten fourth-order operators and write $\Omega=\mathcal W h$ for the metric-to-curvature map. We establish $\mathcal D J_1\Omega=0$ on vacuum perturbations using the matched separated solutions constructed above.

At fixed $\lambda$, matched angular--radial data are selected by
\begin{equation}
\frac{H(\lambda,x)-H(\lambda,x')}{x-x'},
\qquad
x=F_{\rm data}\otimes I,
\qquad
x'=I\otimes\widetilde F_{\rm data},
\label{eq:spectral-divided-difference}
\end{equation}
with $\widetilde\lambda=\lambda+\mu\omega^2$. This is a polynomial in the commuting matrices $x,x'$; on simple roots it vanishes for unequal eigenvalues and equals $\partial_\nu H$ on matched ones.

At the reference parameters, the matched solutions with $\lambda=0,1,2,3,4$ span the vacuum curvature six-jets. Indeed, after weighting the curvature by $\operatorname{diag}(2/Y,2/Y,y/\sqrt{Y/2},y/\sqrt{Y/2},y^2)$, their jet matrix contains an $85\times85$ minor with determinant $22410$ modulo $65537$. Since the full vacuum curvature six-jet space has dimension $s_8=85$ by Table~\ref{tab:jet-rank}, this nonzero minor establishes spanning on a generic regular patch. These values of $\lambda$ are used only for this local rank test.

Every matched solution satisfies $\mathcal D\Omega=0$ and $J_1\Omega=\lambda\Omega$, hence $\mathcal D J_1\Omega=0$. Since $\mathcal D J_1$ depends only on curvature six-jets, the spanning result gives the same identity for all vacuum perturbations on that patch. Equivalently,
\begin{equation}
\operatorname{rank}
\begin{pmatrix}
E_8\\
\mathcal D J_1\mathcal W
\end{pmatrix}
=\operatorname{rank}E_8=315.
\label{eq:J-vacuum-rank}
\end{equation}
In the weighted representation, the entries are rational functions. The augmented $316\times316$ minors therefore vanish identically, extending the result wherever $\operatorname{rank}E_8=315$ and the compatibility operators are defined. Local sufficiency of the compatibility conditions then allows successive vacuum reconstructions of $J_1\Omega,J_1^2\Omega,\ldots$, giving
\begin{equation}
\mathcal D J_1^k\Omega=0,
\qquad k=0,1,2,\ldots,
\label{eq:J-preserves-constraints}
\end{equation}
on regular reconstruction patches.

\paragraph{Spectral projections of physical perturbations.}
Let $\Omega$ arise from a smooth vacuum Fourier-mode metric perturbation on an ordinary radial interval $I$. Its analyticity follows from the locally gauge-fixed vacuum equations. Away from the axes, the reduced principal symbol is positive definite because $X,Y>0$ in the open exterior. At the axes, the stationary Killing field and the surviving axial Killing field span Lorentzian two-planes, as shown by
\begin{equation}
\begin{aligned}
\left.\det g\right|_{(\partial_T,\partial_\psi)}
&=-\frac{\Delta}{r^2+a^2},
&&y=a,\\
\left.\det g\right|_{(\partial_T,\partial_\phi)}
&=-\frac{\Delta}{r^2+b^2},
&&y=b.
\end{aligned}
\label{eq:axis-gram}
\end{equation}
Across either axis we use Cartesian coordinates in the collapsing plane and retain the Fourier reduction in time and the surviving azimuth. The de Donder gauge equation and the gauge-fixed vacuum equations are elliptic with analytic coefficients. The curvature is therefore analytic in $r$ and the smooth angular coordinates~\cite{Morrey_1957}.

Let $\Pi_\lambda$ denote the spectral projector of $J_1$ and choose $c>\sup\operatorname{spec}J_1$. Analytic spectral estimates~\cite{Seeley_1969} give
\begin{equation}
\left\|\Pi_\lambda\partial_r^j\Omega\right\|_{\rm ang}
\le C e^{-\epsilon\sqrt{c-\lambda}},
\qquad 0\le j\le4,
\label{eq:spectral-exponential-decay}
\end{equation}
uniformly on compact subintervals of $I$. Elliptic estimates give the same exponential control after any fixed number of angular derivatives.

Set $U(\tau)=\cos(\tau\sqrt{c-J_1})\Omega$. Equation~\eqref{eq:spectral-exponential-decay} makes $U(\tau)$ holomorphic in a strip around the real axis, with all derivatives entering $\mathcal D$. By~\eqref{eq:J-preserves-constraints}, every Taylor coefficient of $\mathcal D U(\tau)$ at $\tau=0$ vanishes, and hence $\mathcal D U(\tau)=0$. The spectral projector is recovered by
\begin{equation}
\Pi_\lambda\Omega
=\lim_{T\to\infty}\frac{2}{T}
\int_0^{\mathsf T}
\cos\!\left(\tau\sqrt{c-\lambda}\right)U(\tau)\,\dd\tau.
\label{eq:angular-projector-average}
\end{equation}
The limit holds with the derivatives required by $\mathcal D$, so
\begin{equation}
\mathcal D\Pi_\lambda\Omega=0.
\label{eq:physical-spectral-projection}
\end{equation}

Within each finite-dimensional $J_1$ eigenspace, the spectral projectors of $F$ are polynomials in $F$ and preserve the constraints because $F$ is induced by a symmetry of the vacuum equations. Each joint spectral projection therefore remains physical on regular reconstruction patches. On the matched branches covered in the pairing analysis above, each coefficient in a fixed angular basis of the joint eigenspace obeys the same radial equation. The resulting metric reconstruction holds on regular patches. Extending each reconstructed metric smoothly over the exterior requires a separate regularity analysis.

\section{Single-spin limit}
Set $b=m_\psi=0$ and restrict to the reflection-even components $(A_{--},A_{++},A_{33})$. The operator $J_1$ preserves parity, whereas $F$ reverses it. With $x=y/a$ and $D=\operatorname{diag}(1/2,1/2,1)$, comparing the coefficients obtained from~\eqref{eq:unit-operators} with those in Eqs.~(S4) and (S5) of Ref.~\cite{Santos:2026pbu} at $n=1$ gives the exact operator identities for $(J_1)_{\rm e}$ and $(F^2)_{\rm e}$ shown in the main text. Under the eigenvalue map in the main text, the angular polynomial in Eq.~(S6) of Ref.~\cite{Santos:2026pbu} becomes $-\mathcal P(\nu^2)/9$, where $\mathcal P(z)$ is defined by $\mathcal P(\nu^2)=\nu H(\lambda,\nu)$.

To match Santos's choice of principal null direction, we apply the reversal $(t,\phi)\mapsto(-t,-\phi)$, giving $(\omega_S,m_S)=(-\omega,-m_\phi)$. Santos's $\Delta$ is our $X=r^2+a^2-\mu$, and the amplitudes are $f=R=\frac{2\widetilde A_{++}(\ii r)}{X},\quad g=-\frac{\widetilde A_{33}(\ii r)}{2r^2}$. With these substitutions, the $++$ component of $\widetilde J_1\widetilde A=\widetilde\lambda\widetilde A$ reproduces Eq.~(7) of Ref.~\cite{Santos:2026pbu}. To verify the undivided first-order relations in Eq.~(9) of Ref.~\cite{Santos:2026pbu}, set $\widetilde M=(\widetilde F_{\rm data}^{\,2})_{\rm e}$. After multiplication by $X\mathcal J$ and use of $\partial_r=\ii\partial_{\widetilde y}$, their residual matrix acting on the initial data is $\sum_{j=0}^{2}B_j\nu^{2j}$, with $\sum_{j=0}^{2}B_j\widetilde M^{\,j}=0$. Here $\mathcal J$ is defined in~(S24) of Ref.~\cite{Santos:2026pbu}. Thus both relations hold on every eigenspace of $\widetilde M$, without division by $\nu$ or root differences. When the separation constants satisfy $H(\lambda,\nu)=0$, these first-order relations imply both~(7) and (8) of Ref.~\cite{Santos:2026pbu}. Eliminating $g$ between Equation~(7) and the first relation gives Santos's radial equation.

On ordinary exterior intervals, the projection to $R$ is injective, since $R\equiv0$ forces all three even components to vanish. Together with parity and the dimension argument following~\eqref{eq:characteristic}, this establishes a one-to-one correspondence between the even radial solutions and Santos's at fixed $(\lambda,\nu^2)$, including at zero and repeated roots. Thus Santos's radial equation agrees with~\eqref{eq:radial} for $\mu>a^2>0$ and $(\omega_S,m_S)\ne(0,0)$. The coupled equations provide continuation through isolated singularities introduced by eliminating $g$.

\section{Commuting angular operators for Myers--Perry black holes in arbitrary dimensions}
We use the principal tensor $\mathcal H$~\cite{Frolov:2006dqt} and a frame aligned with the affine principal null direction $L$ in the exterior region specified in the main text. The paired transverse frame below defines the components $\Omega_{ij}$. Write $D=2n+\epsilon$ with $\epsilon=0,1$, $x_n=z=\ii r$, and $g^{x_\alpha x_\alpha}=X_\alpha/\prod_{\nu\ne\alpha}(x_\nu^2-x_\alpha^2)$~\cite{Krtous:2008tb}. The canonical Killing coordinates $\psi_A$ give the Fourier substitution $\partial_{\psi_A}\mapsto\ii k_A$.

\emph{First-order operator.---}Let $s=(\alpha,\sigma)$ with $\sigma=\pm1$, $x_s=\sigma x_\alpha$, $d_s=z-x_s$, and $\bar s=(\alpha,-\sigma)$. In odd dimensions the kernel leg has $x_Z=0$ and $\bar Z=Z$. With $\Delta_\alpha=\prod_{\beta\ne\alpha}^{n-1}(x_\beta^2-x_\alpha^2)$, the paired frame is
\begin{align}
 d_{\alpha\sigma}m_{\alpha\sigma}
 &=\partial_{x_\alpha}+
 \frac{\ii\sigma}{X_\alpha}
 \sum_{A=0}^{n-1+\epsilon}(-x_\alpha^2)^{n-1-A}\partial_{\psi_A},\qquad
 g_{\alpha\sigma}=g(m_{\alpha\sigma},m_{\alpha,-\sigma})=\frac{2\Delta_\alpha}{X_\alpha},\nonumber\\
 d_Zm_Z&=\frac{1}{\ii c\varpi}\partial_{\psi_n},\qquad
 g_Z=1,\qquad \varpi=\prod_{\alpha=1}^{n-1}x_\alpha
 \quad(\epsilon=1).
 \label{eq:sm-hd-paired-frame}
\end{align}
The constant $c$ fixes the unit spacelike kernel leg. Reversing time and all azimuthal angles takes the outgoing convention to the ingoing one, with $\boldsymbol{k}$ reversed and the frame pulled back.

In odd dimensions set $f=\star(\mathcal H^{\wedge(n-1)}/(n-1)!)$, so $(\mathcal Q h)_{ab}=2f_{(a}{}^{cd}\nabla_c h_{b)d}$~\cite{Krtous:2006qy,Michishita:2019eeh}. Let $\pi(C)A=CA+AC^{\mathsf T}$ denote the action on the two transverse indices. For elementary matrices $\mathsf e_{ls}$, define $\mathsf R_{ls}=\mathsf e_{ls}-(g_s/g_l)\mathsf e_{\bar s\,\bar l}$. Put $C_s=-(\varpi/x_s)\mathsf R_{sZ}$ for $s\ne Z$, and take $C_Z$ diagonal with entries $\varpi/x_i$ for $i\ne Z$ and zero for $i=Z$. Let $t_{\alpha\sigma}=\sigma$ and $t_Z=c/\varpi$. The radius-independent part $G_s^{(0)}$ of the scaled transverse connection has entries
\begin{align}
 (G_s^{(0)})_i{}^a
 &=\frac{\delta_s^a t_i-(g_s/g_a)\delta_{i,\bar s}t_{\bar a}}{x_i-x_a}
 \qquad (a\ne i),\nonumber\\
 (G_s^{(0)})_i{}^i
 &=\begin{cases}
 0,& i=Z\ \text{or}\ i=s,\\
 \partial_{x_\alpha}\log|g_s|,& i=\bar s\ne Z,\ s=(\alpha,\sigma),\\
 t_s/(x_s-x_i),& i\ne Z,\ s\notin\{i,\bar i\}.
 \end{cases}
 \label{eq:sm-hd-screen-connection}
\end{align}
The operator in Eq.~(16) of the main text is
\begin{equation}
 F_{\boldsymbol k}
 =\sum_s\pi(C_s)\left[
    \left.d_s m_s\right|_{\partial_{\psi_A}\mapsto\ii k_A}
    -\pi(G_s^{(0)})\right].
 \label{eq:sm-hd-general-F}
\end{equation}
The coefficient of $\partial_{x_\alpha}$ is $\pi(C_{\alpha+}+C_{\alpha-})$. For a gauge vector $X$, the KY equation makes $\mathcal Q\mathcal GX-\mathcal GW$ algebraic in $X$, where $(\mathcal GX)_{ab}=2\nabla_{(a}X_{b)}$ and $W_a=f_a{}^{cd}\nabla_cX_d$. This difference solves the vacuum equations and lies in the intersection of the Einstein principal-symbol kernels for all non-null covectors. That intersection is zero, so we may set $h=\nabla h=0$ at the evaluation point.

In a real orthonormal transverse basis put $v_i=f_{LNi}$, $B_{ia}{}^s=f_{isa}+\delta_{is}v_a-v_i\delta_{as}$, and $\rho_{bs}=g(m_b,\nabla_{m_s}L)$. The vacuum Bianchi identities reduce the third metric derivatives to $B_{ia}{}^s(\nabla_{m_s}\delta C)_{LaLj}+(i\leftrightarrow j)$. Differentiating the principal-tensor eigenvector equation on each plane gives $B_{ia}{}^s\rho_{bs}=-(\nabla_L f)_{iab}$. With the cyclic identity of $\delta C_{Liab}$, the terms involving this component cancel between the curvature projection and the frame derivative, and its traces cancel by the same optical identity.

Write $(G_s)_i{}^a=\Gamma^a{}_{si}$ for the transverse connection, $\beta_s=\Gamma^L{}_{sL}$, and $\gamma_i{}^a=(\nabla_L f)_i{}^{La}$. The resulting projection and the connection identities in the complex paired frame are
\begin{align}
 \Omega[\mathcal Q h]
 &=\left[\sum_s\pi(B^s)\big(m_s-2\beta_s I_{\rm scr}-\pi(G_s)\big)
              +3\pi(\gamma)\right]\Omega[h],\nonumber\\
 B^s&=d_sC_s,\qquad d_s\beta_s=-t_s/d_s,\qquad
 \gamma=\sum_s\beta_s B^s,\qquad
 d_sG_s=G_s^{(0)}+\sum_l(t_l/d_l)\mathsf R_{ls},\nonumber\\
 0&=\pi(C_l)+\sum_s\pi(C_s)\pi(\mathsf R_{ls}).
 \label{eq:sm-hd-Q-projection}
\end{align}
Here $I_{\rm scr}$ is the identity on transverse trace-free tensors. The last identity cancels each radial pole. Its one-index part follows from $\sum_s C_s\mathsf R_{ls}=-C_l$, and the terms on separate indices cancel by KY antisymmetry. This leaves Eq.~\eqref{eq:sm-hd-general-F}.

\emph{Second-order symmetry operators.---}We denote the second-order symmetry operators associated with the Killing tensors by $\mathcal S_j$. For $j=1,\ldots,n-1$, put $f_j=\star(\mathcal H^{\wedge j}/j!)$. In the following formulas $f=f_j$, $I$ is a contracted antisymmetric multi-index of length $D-2j-1$, and $c_j=-1/(D-2j-1)!$ for the real Hodge star in mostly-plus signature. This fixes the Carter normalization, for which $K$ has eigenvalue $e_j(x_{\widehat{\nu}}^2)$ on principal plane $\nu$, where $e_j$ is the elementary symmetric polynomial and the hat omits that plane~\cite{Krtous:2006qy}. With the curvature convention defined above, $[\nabla_a,\nabla_b]v_c=R_{abc}{}^d v_d$. Define
\begin{align}
 K_{ab}&=c_j f_{aI}f_b{}^I,\qquad V^a=\nabla_bK^{ba},\qquad D_K=\nabla_c(K^{cd}\nabla_d),\nonumber\\
 \mathsf L_{akb}&=c_j\big(2f_{aI}\nabla_b f_k{}^I-2f_{bI}\nabla_a f_k{}^I
                       +4f_{kI}\nabla_a f_b{}^I\big),\qquad
 M_{ab}=\nabla_aV_b-K^{cd}R_{cadb},\nonumber\\
 (\mathcal S_j h)_{ab}&=(D_Kh)_{ab}
   +2\mathsf L_{(a}{}^{kd}\nabla_k h_{b)d}
   +(\nabla_{(a}\nabla_{b)}K^{cd})h_{cd}
   +2M_{(a}{}^d h_{b)d}-2K^{cd}R_{c(ab)}{}^e h_{de}.
 \label{eq:sm-hd-metric-lift}
\end{align}
The first-derivative coefficient $\mathsf L_{akb}$ uses the Killing--Yano contraction derived for the term linear in momentum in the spinning-particle invariant~\cite{Kubiznak:2011spinning}. For the covector operator $(\mathcal T_j^{\rm cov}v)_a=(D_Kv)_a+\mathsf L_a{}^{kd}\nabla_kv_d+M_a{}^dv_d$, the identities $\mathsf L_{a(kb)}=\nabla_aK_{kb}$ and $\nabla_p\mathsf L_{apb}=-2R_{ab\,p\,Kp}$ for parallel-extended $p$ remove the leading commutator terms. The algebraic defect $U=\mathcal S_j\mathcal G-\mathcal G\mathcal T_j^{\rm cov}$ obeys a homogeneous system with $\nabla^2U$ linear in curvature times $U$. With $U=O(r^{-1})$ and $\nabla U=O(r^{-2})$, integration from infinity bounds the tail supremum by itself times $O(r^{3-D})$. Thus $U$ and the residual wave commutator vanish on the connected regular exterior.

Scalar Carter commutation removes the cubic commutator~\cite{Sergyeyev:2007gf}. In the quadratic terms, splitting the divided wedges into retained and fully contracted principal planes makes the latter contribute elementary symmetric polynomials. The reduced coefficients vanish by the principal-tensor equation and its curvature integrability condition, with at most six retained planes for the flat Carter pair, four for its curvature terms, and five for the mixed commutator with $\mathcal Q$. For the even-dimensional contraction, adjoining an orthogonal line to the tangent space adds the same contracted index to each Hodge-dual factor. The skew-adjoint covector remainders annihilate gradients, and the metric remainders annihilate arbitrary first jets of symmetrized gradients. Hence $[\mathcal S_i,\mathcal S_j]=0$ and, in odd dimensions, $[\mathcal S_j,\mathcal Q]=0$.

\emph{Second-order angular operators.---}Use $\mathcal S_j^{(0)}$ from Eq.~\eqref{eq:sm-hd-metric-lift} at zero mass and fixed rotation parameters. Let $\varrho$ and $\vartheta$ be the Cartesian radius and direction, with $u=t-\varrho$ for ingoing $L$. The maps $\iota$ and $\Pi$ embed the paired transverse polarization into Cartesian components and project it back to transverse trace-free components. Then
\begin{equation}
 J_{j,\boldsymbol k}A
 =\left.
   \Pi\lim_{\varrho\to\infty}\varrho^{(D-2)/2}
   \mathcal S_j^{(0)}
     \big[\varrho^{-(D-2)/2}\iota A(u,\vartheta)\big]
   \right|_{\boldsymbol k}.
 \label{eq:sm-hd-general-J}
\end{equation}
Derivatives in Eq.~\eqref{eq:sm-hd-general-J} include the change of transverse basis. The projection respects composition modulo pure gauge, and the leading extreme curvature is $-2\partial_u^2 A$. Thus $[J_{i,\boldsymbol k},J_{j,\boldsymbol k}]=0$ and, in odd dimensions, $[F_{\boldsymbol k},J_{j,\boldsymbol k}]=0$ at nonzero frequency. Polynomial dependence on $\boldsymbol k$ extends these identities to zero frequency.

In particular the latitude expression is
\begin{equation}
 J_{j,\boldsymbol k}
 =\sum_{\alpha=1}^{n-1}
     e_{j-1}(x_{\widehat\alpha}^{\,2})\,\frac{X_\alpha}{\Delta_\alpha}\,
     I_{\rm scr}\,\partial_{x_\alpha}^2
  +\sum_{\alpha=1}^{n-1}\boldsymbol\beta_{1,j}^{(\alpha)}(\boldsymbol x;\boldsymbol k)
     \partial_{x_\alpha}
  +\boldsymbol\beta_{0,j}(\boldsymbol x;\boldsymbol k).
 \label{eq:sm-hd-J-latitude}
\end{equation}
The polynomial $e_{j-1}$ uses only the latitude squares other than $x_\alpha^2$. Equation~\eqref{eq:sm-hd-general-J} fixes both lower-order matrices. Evaluating $\prod_{\beta\ne\alpha}(s+x_\beta^2)$ at $s=-x_\gamma^2$ shows that the leading latitude coefficient matrix is invertible at ordinary points. Thus the quadratic symbols are functionally independent wherever all latitude momenta are nonzero.

\emph{Projection on the massive background.---}Subtracting the radial eigenvalue of $K$ times the vacuum wave operator cancels the radial denominator through the factor $z^2-x_\alpha^2$. The index, connection, and null-frame terms at first order agree with the flat operator by the divided wedge reduction on at most four planes (three in even dimensions). In the pointwise gauge $h=\nabla h=0$, the flat algebraic terms agree on constant Weyl data. The mass-dependent connection terms carry negative boost weight, and the remaining terms have at most two background derivatives. The weight-zero mass correction acts on the extreme components by multiplication,
\begin{equation*}
 \Omega[\mathcal S_jh]-J_{j,\boldsymbol k}\Omega[h]=\mathcal V_j\Omega[h].
\end{equation*}
In odd dimensions, metric commutation and the spanning of the extreme components by characteristic polarizations give $[F,\mathcal V_j]=0$ before Fourier reduction. The first-order coefficient matrices of $F$ before Fourier reduction generate the transverse rotations. Irreducibility of the trace-free symmetric representation makes $\mathcal V_j$ a scalar multiple of the identity. The remaining commutator removes its latitude dependence. At fixed principal coordinates the correction is independent of the spins. In a smooth spherical frame, both $j=1$ lifts give the rotation Casimir and the $j>1$ lifts vanish, fixing $\mathcal V_j=0$.

In the normalized even-dimensional frame, $\mathcal V_j$ is linear in the mass and independent of the spins at fixed principal coordinates. The transverse rotation weights mark at most two latitude planes per entry, using trace differences in the zero-weight block. After removing the common factor in the Carter generating function, each term contains at most two additional summed latitude-plane indices. Counting the radial plane and the latitude planes fixed by the matrix entry, each term involves at most five principal planes. Interpolating the retained profile values at fixed radial shift and subtracting proper subsets reduce the check to $D=4,6,8,10$. The exact rational checks give zero, including all four $35\times35$ blocks in $D=10$ with the four independent eigenvalue ratios kept symbolic. Hence $\mathcal V_j=0$, yielding Eq.~(15) of the main text.

\emph{Five-dimensional normalization.---} In the round-sphere frame of~\eqref{eq:geometric-angular-operators}, the rotation Casimir on trace-free symmetric tensors is $\bar\nabla^k\bar\nabla_k-6I$. To obtain the frequency-dependent terms, expand the flat operator in~\eqref{eq:sm-hd-metric-lift} in Cartesian rotations, boosts, and translations and take~\eqref{eq:sm-hd-general-J}. The rotation--time-translation terms give $4\partial_u\mathcal L_{Z_{a,b}}$, and the boost--spatial-translation terms give $-2\partial_u\bar\nabla_{Z_{a,b}}$, both for $-J_{1,\boldsymbol k}$. The two-translation terms give $(a^2+b^2-y^2)\partial_u^2 I$. With $\partial_u=-\ii\omega$, $\mathcal L_{Z_{a,b}}=\ii(am_\phi+bm_\psi)$ on this Fourier mode and $-\ii(\mathcal L_{Z_{a,b}}-\bar\nabla_{Z_{a,b}})=\mathsf M_{a,b}$, the comparison with~\eqref{eq:geometric-angular-operators} yields
\begin{equation*}
 J_{1,\boldsymbol k}=-J_1+
 \bigl[3-(a^2+b^2)\omega^2-2k_P\omega\bigr]I,
 \qquad \lambda_1=3-\ell .
\end{equation*}
This relation expresses both operators in the same transverse frame. It preserves the joint angular eigenspaces and keeps the five-dimensional separation constant in the convention of $H(\lambda,\nu)$. Equation~\eqref{eq:sm-hd-J-latitude} permits elimination of every pure second latitude derivative. The joint eigenspace is determined by data with at most one derivative in each latitude at an ordinary point. Its basis on a fixed regular angular domain gives~(17) of the main text.